\documentclass[sigconf]{acmart}
 \usepackage{amsmath,amssymb,amsfonts,bm}
\usepackage{graphicx}
\usepackage{caption}
\usepackage{subfig} 
\usepackage{booktabs} 
\usepackage{textcomp}
\usepackage{xcolor}
\usepackage{float}
\usepackage{array}
\usepackage{tabularx}
\usepackage{array}
\usepackage{balance}
\usepackage{marvosym}
\usepackage{multirow}
\usepackage{array}
\usepackage{enumitem}
\usepackage{booktabs}
\usepackage{algpseudocode}
\usepackage[ruled]{algorithm2e} 
\usepackage{xcolor}
\usepackage{pifont}
\usepackage{url}
\usepackage{algpseudocode}
\usepackage{subfig}
\usepackage[ruled]{algorithm2e}
\copyrightyear{2026}
\acmYear{2026}
\setcopyright{cc}
\setcctype{by}
\acmConference[WSDM '26]{Proceedings of the Nineteenth ACM International Conference on Web Search and Data Mining}{February 22--26, 2026}{Boise, ID, USA}
\acmBooktitle{Proceedings of the Nineteenth ACM International Conference on Web Search and Data Mining (WSDM '26), February 22--26, 2026, Boise, ID, USA}
\acmPrice{}
\acmDOI{10.1145/3773966.3777961}
\acmISBN{979-8-4007-2292-9/2026/02}

\begin{document}

\title{On-Device Large Language Models for Sequential Recommendation}
\author{Xin Xia}
\email{x.xia@uq.edu.au}
\affiliation{
  \institution{The University of Queensland}
  \city{Brisbane}
  \state{QLD}
  \country{Australia}
}

\author{Hongzhi Yin}
\email{h.yin1@uq.edu.au}
\affiliation{
  \institution{The University of Queensland}
  \city{Brisbane}
  \state{QLD}
  \country{Australia}
}

\author{Shane Culpepper}
\authornote{Corresponding author.}
\email{s.culpepper@uq.edu.au}
\affiliation{
  \institution{The University of Queensland}
  \city{Brisbane}
  \state{QLD}
  \country{Australia}
}
\begin{abstract}
On-device recommendation is critical for a number of real-world applications, especially in scenarios that have agreements on execution latency, user privacy, and robust functionality 
when internet connectivity is unstable or even impossible.
While large language models (LLMs) can now provide exceptional capabilities that model user behavior for sequential recommendation tasks, their substantial memory footprint and computational overhead make the deployment on resource-constrained devices a high risk proposition. In this paper, we propose OD-LLM, the first task-adaptive compression framework explicitly designed to provide efficient and accurate on-device deployment of LLMs for sequential recommendation tasks. 
OD-LLM uniquely integrates two complementary compression strategies: a low-rank structural compression algorithm which uses Singular Value Decomposition (SVD) to significantly reduce parameter redundancy in the model, and a novel tokenization normalization technique that better complements the low-rank decomposition process being used.
Additionally, to minimize any potential performance degradation when using higher compression ratios, a novel progressive alignment algorithm is used to iteratively refine the parameters required layerwise in the target model.
Empirical evaluations conducted on sequential recommendation benchmarks show that OD-LLM exhibits no loss in effectiveness when compared to the original recommendation model, when the deployed model size is halved. These promising results demonstrate the efficacy and scalability of OD-LLM, making this novel solution a practical alternative for real-time, on-device solutions wishing to replace expensive, remotely executed LLMs.

\end{abstract}




\maketitle

\section{Introduction}
On-device recommendation systems \cite{yin2024device, yuan2023federated,zhong2025towards,yuan2024hetefedrec} are rapidly gaining importance due to the increasing demand for real-time personalized services, more stringent privacy controls, and frequent network connectivity issues.
Applications ranging from mobile streaming platforms and wearable health devices to smart home environments integrate sequential recommendation models, and there is increasing pressure for recommendation models to operate fully on-device \cite{dhar2021survey, incel2023device}. 
Localized inference capabilities not only reduce latency and enable continuous functionality regardless of internet connectivity, and also address criticisms related to data privacy, by avoiding sensitive information transmission to remote SaaS servers. 
This is particularly crucial for applications that incorporate user-specific behavioral data, such as viewing history, location patterns, or personal profiles, which can be misused if transmitted or stored remotely. 
By keeping both data and inference on-device, these systems reduce the risk of data breaches, are more likely to comply with data protection regulations, and increase user trust. 
As data compliance demands continue to grow, it is imperative to enable advanced recommendation models to be developed that can deliver high-quality predictions entirely from edge devices, without relying on expensive cloud infrastructure and services to function correctly.

Recent advances in large language models (LLMs), such as GPT  \cite{achiam2023gpt} and LLaMA \cite{touvron2023llama}, have significantly enhanced the performance of sequential recommendation tasks through their superior ability to capture intricate user preferences and complex interaction patterns over extended user histories \cite{wu2024survey}.  
By pretraining on vast and diverse corpora, and their impressive abilities to generalize, allow them to process context-rich input prompts and deliver high-quality, semantically aware recommendations. 
As a result, LLMs are now a requirement in many personalized recommendation tasks, and recent efforts explore how to adapt these models to domain-specific scenarios, such as entertainment, e-commerce, and healthcare. 
Motivated by promising preliminary results, recent industry trends are pushing to make LLMs more accessible in local environments, which has led to rapid developments in on-device LLMs that are smaller, faster, and increasingly more efficient variants tailored for edge device deployments \cite{xu2024device}.

In both industry and academia, there has been a surge of interest in making LLMs available entirely on resource constrained devices. 
Tech giants such as Google, Meta, and Apple have already introduced lightweight LLMs for developers to use, e.g., Gemini Nano \cite{team2023gemini}, LLaMA variants \cite{chu2023mobilevlm}, and Apple’s on-device models \cite{zhou2025apple} that run efficiently on iOS mobile devices. 
In parallel, academic efforts have accelerated to create more effective model compression techniques\cite{zhu2024survey} and optimization strategies for resource constrained devices \cite{shekhar2024towards} that can facilitate real-time inference. 
These advances included quantization \cite{lin2024awq, liu2024spinquant}, pruning \cite{ma2023llm, fu2024lazyllm}, low-rank approximation \cite{song2024low}, and knowledge distillation \cite{xu2024survey}, all of which have demonstrated promising preliminary results in general NLP tasks. 
However, despite the growing momentum, a significant gap remains in adapting LLMs to on-device recommendation tasks.
Sequential recommendation \cite{kang2018self, sun2019bert4rec} presents a unique challenge — models must be able to capture fine-grained temporal and contextual dependencies from user interaction histories, which can be highly sensitive to unpredictable results using current compression techniques.
Existing on-device LLM techniques
often prioritize performance improvements for generic NLP benchmarks and overlook key requirements and behavior modeling demands unique to recommendation-based tasks. 
Furthermore, the diversity of edge hardware platforms introduces an additional layer of complexity -- often requiring specialized solutions to ensure consistent performance in heterogeneous environments.

Motivated by these challenges, we propose OD-LLM, the first compression framework explicitly designed to support  efficient on-device deployment of LLMs for sequential recommendation tasks. 
OD-LLM combines complementary strategies to effectively reduce model complexity without sacrificing recommendation effectiveness. 
A key component of OD-LLM is low-rank structural compression  that leverages Singular Value Decomposition (SVD) \cite{abdi2007singular}, which can significantly reduce the redundancy in large models. 
To address numerical instability and precision challenges inherent to transformers, we introduce a novel tokenization normalization technique that increases the stability of low-rank decomposition, preserving subtle sequential behavior signals essential to high-quality recommendations.
Additionally, a progressive alignment algorithm has been created that iteratively refines parameters layerwise in large models in order to minimize performance degradation when using more aggressive compression configurations.

OD-LLM has been rigorously validated in extensive experiments using well-known sequential recommendation benchmarks.
Our results showcase the effectiveness and efficiency of our new approach - no performance loss when the model size is halved. The comprehensive analysis underscore the practical value and scalability of OD-LLM, which is a significant step towards making real-time, privacy-preserving, on-device recommendation systems powered by LLMs a reality. The key contributions of this research are:
\begin{itemize}[topsep=0pt]

\item We identify and address a significant gap in the current research by targeting the underexplored problem of deploying LLMs for sequential recommendation in resource-constrained, on-device environments.

\item We propose OD-LLM, a task-adaptive compression framework that integrates low-rank structural compression, with a novel tokenization normalization technique to complement, and a novel progressive alignment algorithm that preserve recommendation accuracy under aggressive compression configurations. 

\item We perform extensive experiments using benchmark recommendation datasets, and demonstrate that OD-LLM achieves substantial model compression without sacrificing recommendation performance, enabling practical on-device deployment of LLMs, which empower developers with more fine-grained control of efficiency-effectiveness trade-offs in the models they wish to deploy. 
\end{itemize}
\vspace{-9pt}
\section{Related Work}
\subsection{LLM-Based Sequential Recommendation}
Remarkable advancements in recent years on LLMs provide exciting new opportunities to improve state-of-the-art performance of sequential recommendation by leveraging these new natural language understanding capabilities.
LLM-based approaches in this area of research can be divided into two categories: LLM-based Recommendations and LLM-based Enhancements.
Recommendation algorithms \cite{ bao2023tallrec, cui2022m6, yuan2025fellas} now integrate LLMs directly into task solution through fine-tuning, prompting, or in-context learning. 
For example, P5 \cite{geng2022recommendation} provides instruction-based tuning for T5 \cite{colin2020exploring} to unify model training for five different recommendation tasks, while TALLRec \cite{bao2023tallrec} uses two-stage instruction-based tuning  to integrate LLaMA \cite{touvron2023llama} into binary recommendation tasks using a few-shot approach. 
GenRec \cite{hou2024large} further extends this idea by using instruction-based tuning for LLaMA models by using plain text examples to create generative recommendations.
In contrast, enhancement-based approaches \cite{hou2023learning, yuan2023go} use LLMs to extract feature or as feature generators that augment more traditional models.
For example, Hou et al. \cite{hou2023learning} use the top layer an LLM for sequential recommendation to create a mixture-of-expert model (TMoE), which in turn can create transfer learning recommendation models for related tasks.
This line of research leverages LLMs to encode mutual information and enable cross-domain recommendations, or design-based recommendations -- where specific templates guide instruction-based tuning of the model. 
Techniques such as soft prompting and leveraging embeddings produced from LLMs are employed to incorporate rich textual information, such as user reviews or item metadata, into more conventional recommendation models.
Generative strategies are quickly gaining traction, where the key idea is to cast recommendation to a text-generation task \cite{bao2024large}, and then exploit LLM reasoning abilities to create context-aware recommendations. 
However, this can be computationally demanding and high resource requirements of LLMs can limit scalability and use in real-time applications for recommendation-based tasks \cite{kaddour2023challenges, achiam2023gpt}.
These limitations provide strong evidence for the growing need of researchers to create more efficient LLM compression techniques, which would enable broader adoption of these techniques which require large-scale datasets/models in resource-constrained environments.

\begin{figure*}[t]
    \centering
    \includegraphics[width=\textwidth]{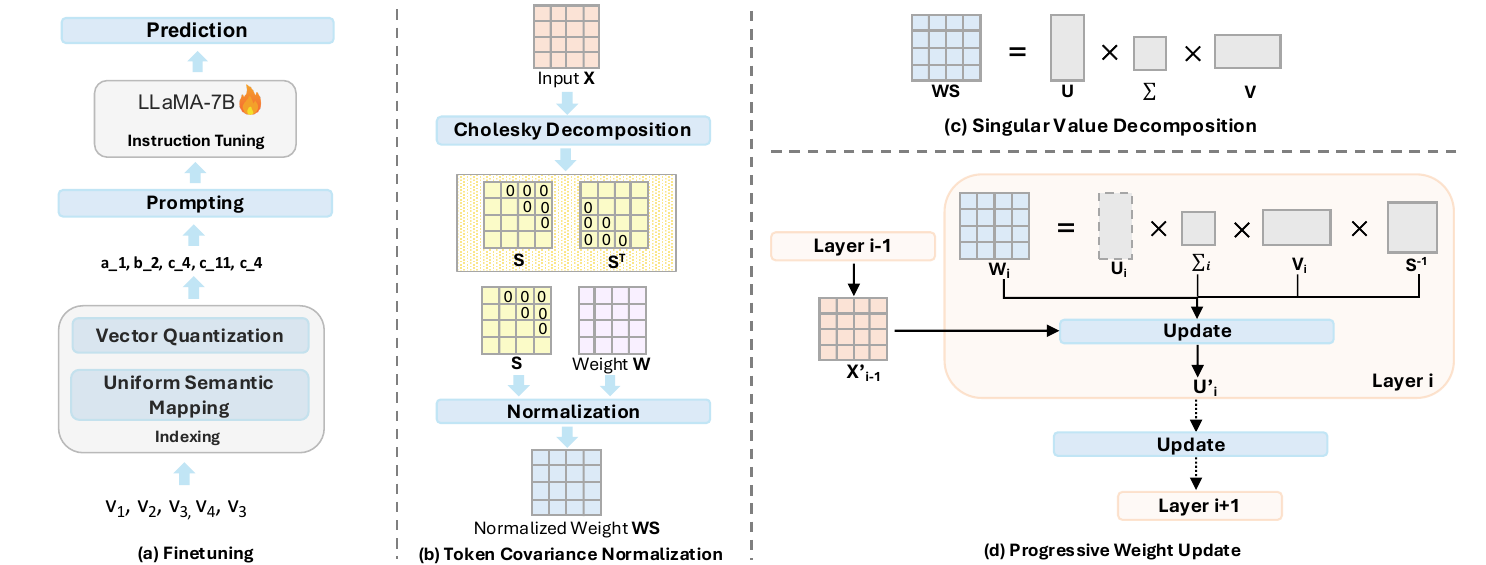}
    \caption{An overview of the proposed LLM-based reecommendation compression technique. }
    \label{figure.1}
    \vspace{-10pt}
\end{figure*}
\vspace{-5pt}
\subsection{LLM-based Compression Techniques} 
The enormous scale of LLMs has increased the attention on model compression in the research community, and is a vital component when deploying these models in resource-constrained environments.
Common compression approaches include quantization \cite{park2022lut, frantar2023sparsegpt}, pruning \cite{sun2023simple, zhang2023dynamic, xia2023flash, molchanov2019importance, ma2023llm}, and knowledge distillation \cite{gu2024minillm, magister2022teaching}.
Quantization tends to reduce parameter precision, as demonstrated by techniques such as Quantization-Aware Training (QAT) \cite{du2024bitdistiller, xu2024onebit} and Post-Training Quantization (PTQ) \cite{park2022lut, frantar2023sparsegpt}, but can lead to a significant decrease in retraining costs, or can lead to unexpected errors in down-stream tasks that rely on rare text features. 
Pruning strategies remove redundant parameters.
Unstructured pruning \cite{sun2023simple, zhang2023dynamic, xia2023flash} can introduce irregular sparsity that requires custom hardware optimization to overcome.
Structured pruning \cite{molchanov2019importance, ma2023llm} can result in substantial performance loss due to coarse-grained removal of important neurons or in the worst case entire layers in the model. 
Knowledge distillation \cite{gu2024minillm, magister2022teaching} uses transfer learning via teacher forcing to induce smaller student models, but the loss of crucial inter-layer connections can limit its applicability in tasks that demand deep contextual understanding.
While all of these approaches are an important step in the right direction for many natural language tasks, their issues described above limit their usefulness in recommendation-based tasks, which require the semantic richness of a model to be preserved, as it is essential in text-driven personalization.
Additionally, many recommender tasks require extensive retraining or specialized dataset calibration, which are not tractable once the model has been processed using these techniques.
These limitations emphasize the growing need for task-adaptive compression techniques that maximize efficiency and minimize performance degradation in real-world applications.
By addressing these challenges, innovative new strategies should provide more control over the trade-offs in efficiency and effectiveness of a model, which would in turn make LLM-based systems more scalable and sustainable than they are today.

\subsection{On-Device Recommendation}
On-device recommendation has recently gained the attention of researchers that are interested in enabling low-latency inference and privacy-preserving personalization.
Prior research used compression in conventional recommendation models through pruning \cite{srinivas2015data}, quantization \cite{gong2014compressing}, and low-rank factorization \cite{oseledets2011tensor}. 
Examples include nearest neighbor search and sequence matching for efficient local prediction \cite{changmai2019device}, tensor-train factorization for compressing embeddings for  next-POI recommendation \cite{wang2020next}.
Other related approaches include TT-Rec which uses caching and improved initialization \cite{yin2021tt}, semi-tensor decomposition to enable flexible session-based recommendation \cite{xia2022device}, elastic embeddings for adaptive device memory constraint matching \cite{chen2021learning}, and pruning with embedding sparsification and local fine-tuning for 10× parameter size reductions \cite{han2021deeprec}.
In parallel, research for on-device LLMs has advanced through quantization, low-rank adaptation, and hardware-aware optimization \cite{qu2025mobile, zheng2025review}, with enterprise solutions such as Google’s Gemini Nano, Meta’s LLaMA variants, and Apple’s mobile-optimized models enabling LLM inference for iOS-based consumer devices. 
However, existing methods are primarily optimized to perform well of general NLP benchmarks and rarely address other important tasks such as sequential recommendation, where compression must preserve fine-grained temporal dependencies and subtle behavioral cues to be effective. 
This research gap motivates our creation of the OD-LLM framework, which customizes LLM compression for the unique requirements found in on-device recommendation.

\section{Methodology}

\subsection{Overview}
In this section, we outline the proposed methodology for compressing LLMs for sequential recommendation systems. 
Our technique is built on an LLM-based recommendation framework, where a pre-trained LLaMA-7B model is fine-tuned with recommendation datasets to effectively capture user-item interaction patterns and collaborative semantics \cite{zheng2024adapting}.
To address the computational and memory challenges of deploying this model on a resource constrained device, we propose a three-step compression framework.
First, we perform Token Covariance Normalization, a preprocessing step that decorrelates and scales token embeddings for stability and efficiency during the compression process.
Next, we apply SVD to factorize the matrix weights and remove singular values below a certain threshold, which reduces the model’s overall rank, while preserving important user-item interaction patterns.
Finally, a layer update step refines the compressed weights by aligning them with updated activations from each preceding layer, ensuring that the representation of the collaborative signals and sequential dependencies are preserved in the neural network.
These three steps ensure that the compressed model retains the ability to provide accurate and context-aware recommendations, even at high compression ratios.

\subsection{Model-based Fine-tuning}

\subsubsection{Sequential Recommendation}
In sequential recommendation, the dataset consists of user-item interaction sequences in a temporal ordering.
We denote the user set as $U$, where $u \in U$, and the item set as $V$, where $i \in V$.
A single user interaction sequence is denoted as $S_u = \{v_1^u, v_2^u, \ldots, v_\ell^u\}$, in chronological order, where $\ell$ represents the length of the sequence.
The objective of sequential recommendation is to predict the next item $v_{\ell+1}^u$ when given any sequence $S_u$.
In LLM-based recommendation models, the model generates recommended items based on the user ID and the user interaction history.

\subsubsection{Fine-tuning}
This work uses LC-Rec \cite{zheng2024adapting} as the basis framework, an open-source platform designed for LLM-based recommendations, which fine-tunes a LLaMa-7B to support recommendation tasks.
Since the focus of this study is on reducing the costs of such a systems, LC-Rec provides a robust fine-tuning methodology, which includes indexing and instruction tuning. 
\par
\textbf{Indexing}: LC-Rec includes a Residual Quantized Variational Autoencoder (RQ-VAE) to encode high-dimensional LLM text embeddings of item descriptions into discrete multi-level index tokens. 
The encoder compresses each item vector into a sequence of coarse-to-fine codewords, where each level has a unique codebook. 
To address indexing conflicts where semantically distinct items are mapped to the same codeword, LC-Rec introduces the notion of Uniform Semantic Mapping. 
This enforces a uniform semantic distribution across codewords using an optimal transport objective using the Sinkhorn–Knopp algorithm, which ensures each index corresponds to a balanced and semantically distinct cluster.
Additionally, the final codebook includes a two-stage redistribution mechanism that relocates conflicting items to semantically consistent positions to further preserve index-semantic fidelity. 
\par

\textbf{Instruction Tuning}: The alignment stage ensures that the LLM understands and predicts discrete index tokens from a recommendation context. 
First, \textit{Sequential Item Prediction} is used to train the model to predict the index of the next item in the sequence, given a sequence of historical item indices. 
Then, \textit{Explicit Index–Language Alignment} pairs index sequences with their natural-language item descriptions bidirectionally, enabling translation between mutual symbolic indices and the semantic text. 
Finally, \textit{Implicit Recommendation-Oriented Alignment} introduces three auxiliary tasks: (a) \textit{Asymmetric Item Prediction}, where input and output histories differ in length to mimic cold-start and partial-history settings; (b) \textit{Item Prediction Based on User Intention}, where explicit intention descriptions are provided to guide the predictions; and (c) \textit{Personalized Preference Inference}, where user profiles are included to personalize predictions for individuals. 
Together, these tasks can enrich how the indices map to nuanced recommendation contexts in the model. 
\par

\textbf{Fine-Tuning Objectives}: Joint training optimizes the RQ-VAE index and the LLM alignment module.
RQ-VAE is trained using a combined reconstruction loss:

\begin{equation}
\mathcal{L}_{\mathrm{recon}} = \| \mathbf{x} - \hat{\mathbf{x}} \|_2^2,
\end{equation}

The RQ loss is:
\begin{equation}
\mathcal{L}_{\mathrm{RQ}} = \sum_{l=1}^{L} \left( \| \mathrm{sg}[\mathbf{z}_l] - \mathbf{e}_l \|_2^2 + \beta \| \mathbf{z}_l - \mathrm{sg}[\mathbf{e}_l] \|_2^2 \right),
\end{equation}
and $\mathbf{z}_l$ is the latent vector at codebook level $l$, $\mathbf{e}_l$ is the corresponding codeword, $\mathrm{sg}[\cdot]$ is the stop-gradient operator, and $\beta$ controls the {\em commitment strength}. 
The optimal transport uniformity loss is:
\begin{equation}
\mathcal{L}_{\mathrm{ot}} = \langle \mathbf{P}^*, \mathbf{C} \rangle,
\end{equation}
where $\mathbf{P}^*$ is the optimal transport plan computed with the Sinkhorn--Knopp algorithm, and $\mathbf{C}$ is the semantic distance matrix between codewords.

The alignment module is trained using a multi-task objective function combining sequential item prediction loss $\mathcal{L}_{\mathrm{seq}}$, explicit index--language alignment loss $\mathcal{L}_{\mathrm{align}}$, and implicit recommendation-oriented task losses $\mathcal{L}_{\mathrm{asym}}$, $\mathcal{L}_{\mathrm{intent}}$, and $\mathcal{L}_{\mathrm{pref}}$.
Each of these losses are a cross-entropy loss over the predicted token sequence:

\begin{equation}
\mathcal{L}_{\mathrm{task}} = - \sum_{t=1}^T \log p_\theta(y_t \mid y_{<t}, \mathbf{x}),
\end{equation}
where $y_t$ is the target token at position $t$, $\mathbf{x}$ is the input context, and $\theta$ are the model parameters.

The overall training loss is:
\begin{equation}
\begin{split}
\mathcal{L} = & \ \lambda_{\mathrm{recon}} \mathcal{L}_{\mathrm{recon}} + \lambda_{\mathrm{RQ}} \mathcal{L}_{\mathrm{RQ}} 
+ \lambda_{\mathrm{ot}} \mathcal{L}_{\mathrm{ot}} + \lambda_{\mathrm{seq}} \mathcal{L}_{\mathrm{seq}}
 + \lambda_{\mathrm{align}} \mathcal{L}_{\mathrm{align}} \\
&+ \lambda_{\mathrm{asym}} \mathcal{L}_{\mathrm{asym}} 
+ \lambda_{\mathrm{intent}} \mathcal{L}_{\mathrm{intent}} + \lambda_{\mathrm{pref}} \mathcal{L}_{\mathrm{pref}}.
\end{split}
\end{equation}
Optimization uses AdamW with a learning rate of $5 \times 10^{-5}$, cosine warmup scheduling, a batch size of 128 with gradient accumulation, and four training epochs. 
In each epoch, every sample is formatted with a randomly selected instruction template to enhance task generalization.

\subsection{Token Covariance Normalization}
In sequential recommendation, token embeddings derived from user-item interaction sequences often exhibit scale variance and occasionally exhibit dimensional correlations, even after fine-tuning a large language model.
These inconsistencies can reduce the efficiency and stability of SVD during compression.
So, no normalization of the correlated dimensions in the embeddings can lead to large, redundant singular values, making it difficult for SVD to distinguish between meaningful features and noise.
Also, scale variance across dimensions increases the risk of certain features dominating the compression decisions, resulting in the loss of the most critical collaborative and sequential patterns.
To address this problem, we apply token covariance normalization using the input activation in each layer.
Before introducing how the technique works, we first provide two necessary definitions:

\begin{definition}Covariance Matrix Transformation.
Given a matrix $\mathbf{X} \in \mathbb{R}^{d \times n}$, the covariance matrix $\mathbf{C}$ is computed as:
\begin{equation}
\mathbf{C} =  \mathbf{X} \mathbf{X}^T,
\end{equation}
where $\mathbf{C} \in \mathbb{R}^{d \times d}$ is a symmetric positive-semidefinite matrix. 
\end{definition}

\begin{definition}Cholesky Decomposition \cite{meyer2023matrix}.
The Cholesky decomposition / Cholesky factorization is decomposing a Hermitian, positive-definite matrix into the product of a lower triangular matrix and its conjugate transpose.
For a matrix $\mathbf{C} \in \mathbb{R}^{d \times d}$, this is defined as:
\begin{equation}
\mathbf{C} = \mathbf{S} \mathbf{S}^T,
\end{equation}
\noindent where $\mathbf{S} \in \mathbb{R}^{d \times d}$ is a lower triangular matrix with positive diagonal entries. 
\end{definition}

Given that $\mathbf{W}$ is the weight matrix in the original LLM, $\mathbf{X}$ is the activation of $\mathbf{W}$ given an input.
The goal of the normalization is to transform the input into a new matrix, such that the covariance matrix becomes the identity matrix $\mathbf{I}$. 
First the covariance matrix of $\mathbf{X}$, i.e. $\mathbf{C}$ is generated:

\begin{equation}
\mathbf{C} =  \mathbf{X} \mathbf{X}^T,
\end{equation}

\noindent and then $\mathbf{C}$ is decomposed:

\begin{equation}
\mathbf{C} = \mathbf{S} \mathbf{S}^T.
\end{equation}

The transformed $\mathbf{X}$ is referred to as $\tilde{\mathbf{X}}$, where $\tilde{\mathbf{X}} =\mathbf{S}^{-1} \mathbf{X} $. 
Therefore, the covariance matrix of $\tilde{\mathbf{X}}$ is:

\begin{equation}
\tilde{\mathbf{C}} = \tilde{\mathbf{X}} \tilde{\mathbf{X}}^T = \mathbf{S}^{-1} \mathbf{X} \mathbf{X}^T (\mathbf{S}^{-1})^{T} = \mathbf{S}^{-1} \mathbf{C} (\mathbf{S}^{-1})^{T} = \mathbf{I}.
\end{equation}

When the covariance matrix of $\tilde{\mathbf{X}}$ is transformed to the identity matrix $\mathbf{I}$, the dimensions of $\tilde{\mathbf{X}}$ are decorrelated, independent, and have equal unit variance.
This ensures that no single dimension can dominate the representation, and simplifies the feature space as the dimensions must be orthogonal.
The transformation improves the numerical stability and ensures that downstream operations such as SVD can operate more efficiently, as singular values directly reflect the importance of uncorrelated features, allowing increased reduction.
In sequential recommendation systems, this process removes redundant correlations, emphasizes unique patterns, and preserves critical collaborative and sequential signals, ensuring efficient compression and more robust performance of the compressed model.

\par

Cholesky decomposition is used in our approach because it is  computationally efficient, increases numerical stability, and is suitable for symmetric positive semi-definite matrices such as $\mathbf{X}\mathbf{X}^T$.
Unlike other methods, such as eigenvalue decomposition \cite{abdi2007eigen}, QR decomposition \cite{gander1980algorithms}, or Principal Component Analysis (PCA) \cite{abdi2010principal}, which are either not efficient or fail to remove unnecessary information, Cholesky decomposition directly transforms the embeddings of $\mathbf{S}^{-1}$, and has more desirable properties for token normalization with minimal computational effort.
When used as a preprocessing step to improve SVD-based compression, Cholesky decomposition offers the the best balance between simplicity and performance, and is the most efficient choice for our problem.

Token covariance normalization decorrelates the dimensionality and scales embeddings uniformly, ensuring that SVD retains the most important features when removing redundancy.
This preprocessing step is essential in minimizing the compression loss, and retains the representational power of the original embeddings, enabling the compressed model to provide higher-quality recommendation performance even under strict resource constraints.

\begin{table}[t]
\centering
\small
\begin{tabular}{lccccc}
\hline
\textbf{Datasets} & \textbf{\#Users} & \textbf{\#Items} & \textbf{\#Inter} & \textbf{Sparsity} & \textbf{Avg. \textit{Length}} \\
\hline
Instruments & 24,773 & 9,923 & 206,153 & 99.92\% & 8.32 \\
Arts        & 45,142 & 20,957 & 390,832 & 99.96\% & 8.66 \\
Games       & 50,547 & 16,860 & 452,989 & 99.95\% & 8.96 \\
\hline
\end{tabular}
\caption{Important properties of the datasets used in our experiments.}
\label{table 1}
\vspace{-2em}
\end{table}

    \begin{table*}[t]
	\begin{center}
		\setlength{\tabcolsep}{1mm}{
		{
		{
			\begin{tabular}{*{13}{c}}
				\toprule
				\multirow{2}{*}{Method} &
				\multicolumn{4}{c}{Instruments} & \multicolumn{4}{c}{Games} & \multicolumn{4}{c}{Arts}\cr
				\cmidrule(lr){2-5}\cmidrule(lr){6-9}\cmidrule(lr){10-13} & HR@5 & NDCG@5 & HR@10 & NDCG@10 & HR@5 & NDCG@5 & HR@10 & NDCG@10   & HR@5 & NDCG@5 & HR@10 & NDCG@10\\ \hline		
							
				Caser  & 0.0543  & 0.0355 &0.0710& 0.0409& 0.0367 &0.0227 &0.0617 &0.0307&0.0379 &0.0262& 0.0541& 0.0313  \\
				
				HGN &0.0813&0.0668& 0.1048&0.0744 &0.0517& 0.0333&0.0856&0.0442&0.0622&0.0462& 0.0875& 0.0544\\
				
				GRU4Rec&0.0821&0.0698&0.1031&0.0765&0.0586& 0.0381&0.0964&0.0502&0.0749&0.0590 &0.0964 & 0.0659\\
				
				Bert4Rec&0.0671&0.0560&0.0822&0.0608& 0.0482&0.0311 &0.0763&0.0401&0.0559&0.0451&0.0713     &0.0500\\
                   
                    FDSA&0.0834&0.0681&0.1046&0.0750&0.0644&0.0404& 0.1041&0.0531&0.0734&0.0595& 0.0933& 0.0660\\
				SASRec&0.0751&0.0627&0.0947&0.0690&0.0581 &0.0365 &0.0940&0.0481&0.0757 & 0.0508 &0.1016     & 0.0592\\
                S3-Rec&0.0863 &0.0626&0.1136& 0.0714&0.0606&0.0364&0.1002&0.0491&0.0767 & 0.0521& 0.1051& 0.0612\\
                TIGER&0.0863&0.0738&0.1064&0.0803&0.0599&0.0392 & 0.0939&0.0501&0.0788& 0.0631& 0.1012 & 0.0703\\
                LC-Rec& \underline{0.0997}& \underline{0.0852}& 0.1217&0.0923 &\underline{0.0876}&0.0635 &0.1252&\underline{0.0757}& 0.1007& 0.0824& 0.1251& 0.0902\\
                OD-LLM&  0.0993& 0.0851 & \textbf{0.1219} &\textbf{0.0923} & 0.0838 &\textbf{0.0644}& \textbf{0.1264} & 0.0748& \textbf{0.1173} &\textbf{0.1000} & \textbf{0.1434} & \textbf{0.1084}\\\hline
                
		\end{tabular}}}}		
	\end{center}
	\caption{A comprehensive performance comparison of all methods.}
	\label{Table 2}
	\vspace{-20pt}	
\end{table*}
\subsection{Singular Value Decomposition}
Token covariance normalization ensures that the token embeddings are decorrelated or standardized by transforming the covariance matrix into an identity matrix.
This preprocessing step simplifies the structure of the embeddings, and ensures that the weight matrix $\mathbf{W}$ can be used on the normalized embeddings.
To use the normalized embeddings, OD-LLM applies SVD to the product $\mathbf{WS}$, where $\mathbf{S}$ is the Cholesky factor derived from the covariance matrix of the original embeddings.
This approach ensures that SVD operates on the normalized feature space, which produces more efficient compression using a direct and interpretable mapping between singular values and the compression loss. 
SVD on $\mathbf{WS}$ produces the decomposed matrices:

\begin{equation}
\mathbf{WS} = \mathbf{U} \mathbf{\Sigma} \mathbf{V}^T,
\end{equation}

\noindent where $\mathbf{W} \in \mathbb{R}^{M \times N}$ and $\mathbf{S} \in \mathbb{R}^{N \times N}$; $\mathbf{U} = [\mathbf{u}_1, \mathbf{u}_2, \mathbf{u}_3, \ldots, \mathbf{u}_r]$, $\mathbf{\Sigma}$ = diag$\left(\mathbf{\sigma}_1, \mathbf{\sigma}_2, \mathbf{\sigma}_3, \ldots, \mathbf{\sigma}_r\right)$, and $\mathbf{V} = [\mathbf{v}_1, \mathbf{v}_2, \mathbf{v}_3, \ldots, \mathbf{v}_r]$; $\mathbf{U}$ and $\mathbf{V}$ are the orthogonal matrices and $\mathbf{\Sigma}$ is a diagonal matrix containing the singular values $\mathbf{\sigma}_1 \geq \mathbf{\sigma}_2 \geq \mathbf{\sigma}_3 \geq \ldots\geq \mathbf{\sigma}_r$; and $r < \min(M, N)$.
To reduce the rank of $\mathbf{WS}$, OD-LLM truncates smaller singular values in $\mathbf{\Sigma}$ below a certain threshold, and reconstructs the compressed weight matrix as:

\begin{equation}
\mathbf{W}' = \mathbf{U} \text{Trunc}(\mathbf{\Sigma}) \mathbf{V}^T \mathbf{S}^{-1}.
\end{equation}

The number of truncated values, i.e. the rank of $\text{Trunc}(\mathbf{\Sigma})$ can be computed as:

\begin{equation}
rank = \left\lfloor \frac{M \cdot N \cdot CR}{M + N} \right\rfloor,
\end{equation}

where $CR \in (0, 1)$ represents the compression ratio.

Reconstruction ensures that the compressed matrix is properly aligned with the original embedding space and preserves the most important components of $\mathbf{W}$. 
So, the compression loss can be defined as:

\begin{equation}
L = \| (\mathbf{W} - \mathbf{W}') \mathbf{X} \|_F,
\end{equation}

\noindent where $\mathbf{X}$ is the activation function of $\mathbf{W}$ given an input, and $L$ is the compression loss in the form of the Frobenius loss.
A critical advantage of our approach is the direct mapping between the compression loss and truncated singular values \cite{wang2024svd}.
That is, when only one singular value \( \mathbf{\sigma}_r \) is truncated, the compression loss is equal to the truncated singular value:

\begin{equation}
L_r = \mathbf{\sigma}_r.
\end{equation}

When there is more than one singular values \( \mathbf{\sigma}_{rank+1}, \mathbf{\sigma}_{rank+2}, \ldots, \mathbf{\sigma}_{r} \) truncated, the squared compression loss of the truncated multiple singular values is equal to the sum of their squares:

\begin{equation}
L = \sum_{i=rank+1}^r \mathbf{\sigma}_i^2,
\end{equation}

\noindent where $rank$ is the number of retained singular values.
So, truncating the smallest singular values results in less compression loss.
By truncating the smallest singular values, the compression loss is minimized but still retains the most important features from the original weight matrix.
This ensures that compression is both efficient and preseves state.
The integration of the previous normalization and SVD ensures minimal compression loss and maximal efficiency in OD-LLM.

\par

Below, we provide the proof of why the proposed normalization technique can provide a direct mapping between singular values and the compression loss.

\begin{lemma}
\label{lemma:label}
The Frobenius norm of a matrix \( A \) of dimension \( m \times n \) can be expressed as the square root of the trace of the Gram matrix:
\begin{equation}
\| A \|_F = \sqrt{\sum_{j=1}^n \sum_{i=1}^m |a_{ij}|^2} = \sqrt{\text{trace}(A^T A)}.
\end{equation}
\end{lemma}

Using Lemma 3.3, we can derive a compression loss \( L_i \) when truncating the \( i \)-th singular value of \( \mathbf{S}^{-1} \mathbf{X} \) to reduce the rank:

\begin{align}
L_i &= \| (\mathbf{W} - \mathbf{W}') \mathbf{X} \|_F \\
    &= \| \mathbf{\sigma}_i \mathbf{u}_i \mathbf{v}_i^T \mathbf{S}^{-1} \mathbf{X} \|_F \\
    &= \mathbf{\sigma}_i \sqrt{\text{trace}(\mathbf{u}_i \mathbf{v}_i^T \mathbf{S}^{-1} \mathbf{X} \mathbf{X}^T (\mathbf{S}^{-1})^T \mathbf{v}_i \mathbf{u}_i^T)}.
\end{align}

Since $\mathbf{U} = [\mathbf{u}_1, \mathbf{u}_2, \mathbf{u}_3, ..., \mathbf{u}_r]$ and $\mathbf{V} = [\mathbf{v}_1, \mathbf{v}_2, \mathbf{v}_3, \ldots, \mathbf{v}_r]$ are orthogonal matrices, we have:
\[
\mathbf{v}_i^T \mathbf{v}_i = \mathbf{u}_i^T \mathbf{u}_i = I, \quad \mathbf{v}_i^T \mathbf{v}_j = \mathbf{u}_i^T \mathbf{u}_j = 0, \quad \forall i \neq j.
\]
Moreover:
\[
\text{trace}(\mathbf{v}_i \mathbf{v}_i^T) = \text{trace}(\mathbf{u}_i \mathbf{u}_i^T) = 1.
\]
Since the covariance normalization matrix \( \mathbf{S} \) is the Cholesky decomposition of \( \mathbf{X}\mathbf{X}^T \), we have \( \mathbf{S}\mathbf{S}^T = \mathbf{X}\mathbf{X}^T \). Using Equation (15).
This can be simplified to:
\begin{align}
L_i &= \| \mathbf{\sigma}_i \mathbf{u}_i \mathbf{v}_i^T \mathbf{S}^{-1} \mathbf{X} \|_F \\
    &= \mathbf{\sigma}_i \sqrt{\text{trace}(\mathbf{u}_i \mathbf{v}_i^T \mathbf{S}^{-1} \mathbf{X} \mathbf{X}^T (\mathbf{S}^{-1})^T \mathbf{v}_i \mathbf{u}_i^T)} \\
    &= \mathbf{\sigma}_i \sqrt{\text{trace}(\mathbf{u}_i \mathbf{u}_i^T)} \\
    &= \mathbf{\sigma}_i;
\end{align}

Thus, the loss \( L_i \) incurred by truncating \( \mathbf{\sigma}_i \) is equal to the singular value \( \mathbf{\sigma}_i \) itself, which proves Equation (11) is true.


If we truncate multiple singular values \( \mathbf{\sigma}_{m+1}, \mathbf{\sigma}_{m+2}, \dots, \mathbf{\sigma}_r \) in \( \mathbf{\Sigma} \), the squared compression loss \( L^2 \) becomes:

\begin{equation}
L^2 = \left\| \sum_{i=m+1}^r \sigma_i \mathbf{u}_i \mathbf{v}_i^T \mathbf{S}^{-1} \mathbf{X} \right\|_F^2,
\end{equation}

which can be expanded to:
\begin{equation}
L^2 = \sum_{i=m+1}^r \sum_{j=m+1}^r \sigma_i \sigma_j \text{trace}(\mathbf{u}_i \mathbf{v}_i^T \mathbf{S}^{-1} \mathbf{X} \mathbf{X}^T (\mathbf{S}^{-1})^T \mathbf{v}_j \mathbf{u}_j^T).
\end{equation}

Since the orthogonal properties of \( \mathbf{u}_i \) and \( \mathbf{v}_i \) ensure \( \text{trace}(\mathbf{u}_i \mathbf{v}_i^T \mathbf{u}_j \mathbf{v}_j^T)\\=0 \) for \( i \neq j \), this can then be simplified to:

\begin{equation}
L^2 = \sum_{i=m+1}^r \sigma_i^2 \text{trace}(\mathbf{u}_i \mathbf{u}_i^T) = \sum_{i=m+1}^r \sigma_i^2.
\end{equation}

Thus, the squared loss \( L^2 \) equals the sum of the squared singular values, which is consistent with Equation (12).

\subsection{Progressive Weight Matrix Updates}
When using SVD to compress a weight matrix, truncating smaller singular values can lead to discrepancies between the original and the compressed model. 
Specifically, as the compression ratio increases, the compressed weight matrix $\mathbf{W}'$ deviates from the original weight matrix $\mathbf{W}$ more, causing the activations produced to differ from those produced by $\mathbf{W}$.
To address this problem, we selectively update the weight matrix $\mathbf{W}$ layerwise, ensuring that the compressed model retains the ability to generate activations which are similar to the original model, thereby minimizing the performance loss introduced by SVD \cite{wang2024svd}.
So, the key innovation is to refine the compressed model by updating each layer independently, using the activations generated in the previous layer.
Rather than globally updating the entire weight matrix after compression, which may lead to lower quality compressed models, we propose a more principled update process.
The weight matrix in each layer is adjusted to reduce the discrepancy between the activations produced by the original and the compressed model. 
The update is performed only on the matrix $\mathbf{U}_i$ for each layer $i$, which represents the left singular vectors of the decomposed weight matrix.
The remaining components, such as $\mathbf{V}_i$ and the truncated singular values in $\mathbf{\Sigma}_i$, are fixed to preserve the low-rank structure of the model.
When given the same calibration data as input, the compressed weight matrix $\mathbf{W}'$ generates a new activation $\mathbf{X}'$ that differs from $\mathbf{X}$, which is generated from the original weight matrix $\mathbf{W}$.
$U_i$ is updated as follows:
\begin{equation}
\mathbf{U}'_i = \mathbf{W}_i \mathbf{X}'_{i-1} \mathbf{D}^T (\mathbf{D} \mathbf{D}^T)^{-1}, \\
\quad \mathbf{D} = \text{Trunc}(\mathbf{\Sigma})_i \mathbf{V}_i^T \mathbf{S}_i^{-1} \mathbf{X}'_{i-1};
\end{equation}

\noindent where $\text{Trunc}(\mathbf{\Sigma})_i$ is the truncated singular value matrix for layer $i$. 
After each progressive layer update, we expect improved performance even at higher compression ratios.

\section{Experiments}

\subsection{Experimental Settings}
\subsubsection{Datasets}
Following previous work \cite{zheng2024adapting}, we evaluate our method using three real-world recommender systems datasets from Amazon review data \footnote{\url{http://jmcauley.ucsd.edu/data/amazon/}}: \textit{Instruments},  \textit{Games} and \textit{Arts}. 
Each item in the dataset is associated with a title and a description.
For any sequence $S_u = \{v_1^u, v_2^u, \ldots, v_\ell^u\}$, the last two interacting items, $v_\ell^u, v_{\ell-1}^u$, are used as the validation and test set respectively. 
A summary of important data statistics is presented in Table \ref{table 1}.

\subsubsection{Baselines}
We first compare our method with a few representative traditional sequential recommendation models and then compare our results using a few small language models.
\par
\begin{itemize}[leftmargin=*]
\item \textbf{Caser} \cite{tang2018personalized} is a novel top-N sequential recommendation system that models recent actions as an `image of time and latent dimensions, and learns sequential patterns using convolutional filters to capture point-level and union-level sequential patterns, skipwise behaviors, and long term user preferences.
\item \textbf{HGN} \cite{ma2019hierarchical} is a hierarchical gating network (HGN), ingegrated with Bayesian Personalized Ranking (BPR) to capture both the long-term and short-term user interests.
\item \textbf{GRU4Rec} \cite{hidasi2015session} proposes an RNN-based approach to model the whole sequence of session-based recommendations.
\item \textbf{Bert4Rec} \cite{sun2019bert4rec} uses deep bidirectional self-attention to model user behavior sequences to learn a bidirectional representation model, and makes recommendations by allowing each item in user behavior history to combine information bidirectionally.
\item \textbf{FDSA} \cite{zhang2019feature} first integrates various heterogeneous features for items into feature sequences and separates self-attention blocks based on item-level sequences and feature-level sequences to model the item transition patterns and the feature transition patterns.
\item \textbf{SASRec} \cite{kang2018self} is a Transformer-based sequential recommendation model that captures long-term semantics and makes predictions using relatively few input actions.
\item \textbf{S3-Rec} \cite{zhou2020s3} employs mutual information maximization to pre-train a self-supervised sequential recommendation model, capturing the relationships between items and the associated attributes.
\item \textbf{TIGER} \cite{rajput2023recommender} adopts a generative retrieval approach to sequential recommendation by introducing semantic IDs to uniquely represent items. 
As the authors have not released the official source code, we re-implement the model using Transformers-3, based on the implementation details outlined in the paper.
\item \textbf{LC-Rec} \cite{zheng2024adapting}: we compare our compressed model using the original LC-Rec model.
\end{itemize}

\subsubsection{Evaluation Metrics}
Since our method is designed for sequential recommendation tasks, we use Hit Ratio and NDCG (Normalized Discounted Cumulative Gain) to evaluate the recommendation results of the top-5 and top-10 results.
The Hit Ratio measures the ratio of correct items predicted and NDCG uses a weighted gain function based on the rank position of the relevant items in the recommendation list.

\subsubsection{Training Settings}
For the LLM-based baselines, we also use a similar fine-tuning strategy for our recommendation datasets.
When training the LLaMA model, we follow the settings from \cite{zheng2024adapting}. For the compression setting, the compression ratio is set to 0.5; the number of calibration samples is 256; the model sequence length is set to 200; the evaluation batch size is 1 and the number of beans is 20. The whole method is done on a single A40 GPU.

\begin{figure}[t]
    \centering
    \includegraphics[width=0.5\textwidth]{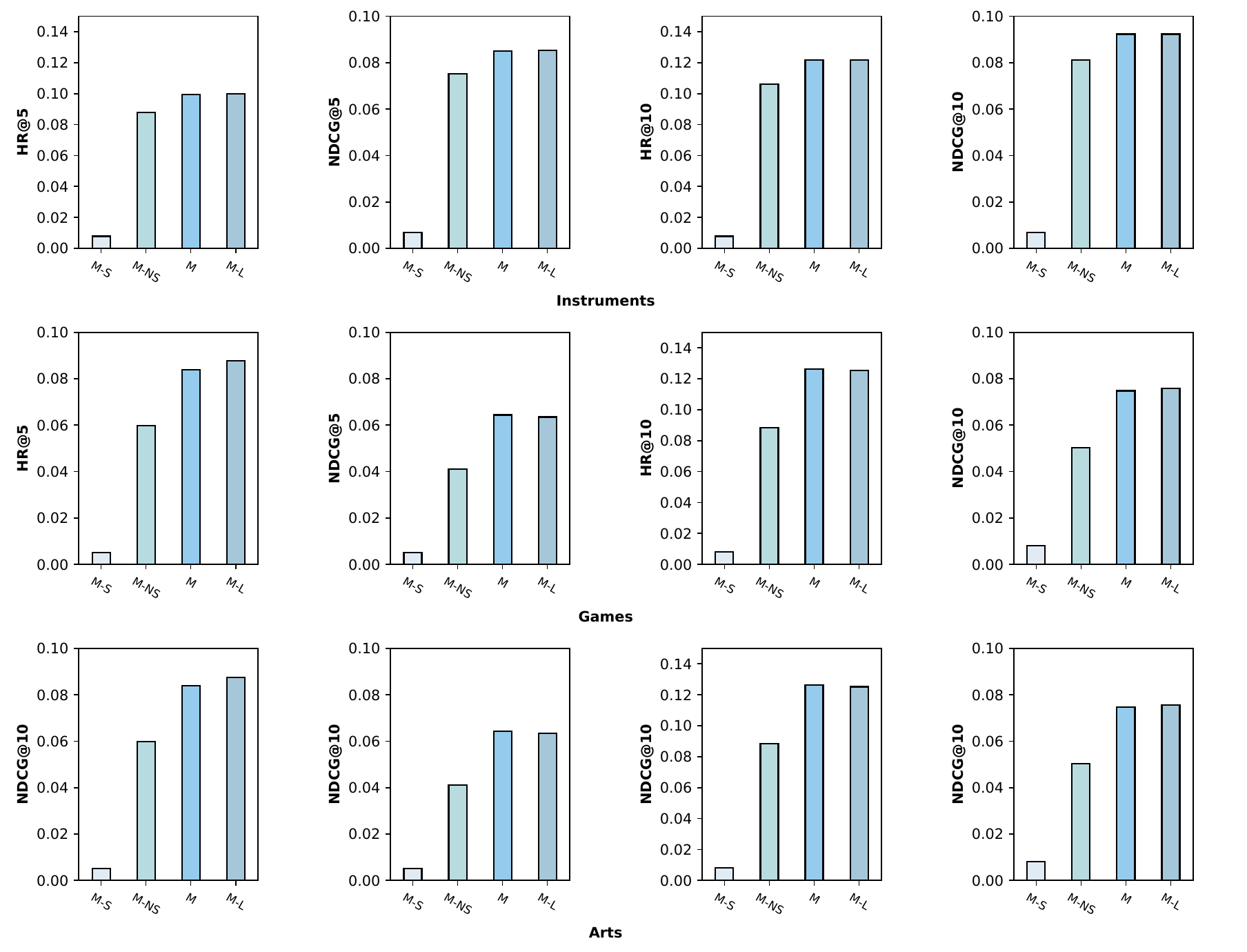}
    \caption{Ablation Study.}
    \label{figure.2}
    \vspace{-10pt}
\end{figure}
\subsection{Overall Performance}

To validate the effectiveness of our proposed method, we conduct experiments using three datasets, and compare the performance against traditional sequential recommendation baselines and also the uncompressed backbone model LC-Rec.
For simplicity, our method represents the model where each weight matrix has a 50\% compression ratio. The results are shown in Table \ref{Table 2}.

Traditional sequential models such as SASRec, GRU4Rec, and HGN are reasonably effective, with GRU4Rec being the best performing baseline for NDCG, indicating better ranking capabilities. 
Bert4Rec and Caser perform poorly in most settings, suggesting that their architectural choices may be less effective at capturing complex user-item interactions. 
Among all baselines, LC-Rec demonstrates the best overall performance across the three datasets, achieving the highest scores for all evaluation metrics considered, which highlights its capacity to integrate collaborative semantics into sequential recommendation. 
Across all three datasets, OD-LLM performs well, or even surpasses, the strongest baseline (LC-Rec). 
On the \textit{Instruments} dataset, the two best methods are essentially indistinguishable for the highest ranks, with OD-LLM having a modest advantage at deeper cutoffs. 
In the \textit{Games} dataset, the relative ranking patterns differ: OD-LLM demonstrates stronger performance at deeper ranks, whereas LC-Rec shows a slight advantage in the highest-ranked results.
In the \textit{Arts} dataset, OD-LLM consistently outperforms LC-Rec across all evaluation metrics, achieving substantial relative improvements.
Recall that LC-Rec is the uncompressed model, whereas OD-LLM can achieve these results using roughly half the model size to deliver similar or even better ranking quality with a significantly smaller memory footprint.

\begin{figure}[t]
    \centering
    \includegraphics[width=0.5\textwidth]{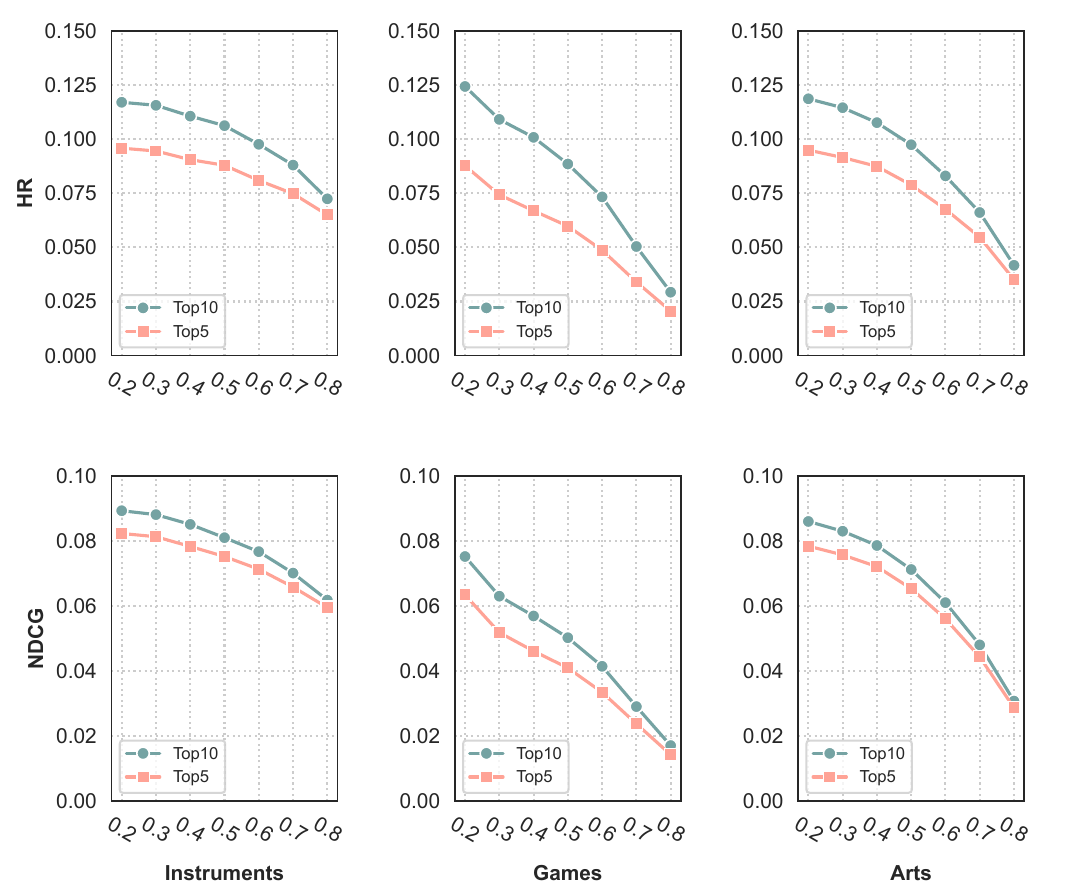}
    \caption{Impact of Compression Ratio.}
    \label{figure.3}
    \vspace{-10pt}
\end{figure}
\begin{table}[ht]
    \centering
    
    \label{tab:quant_all}

    \begin{tabular}{c|c|c|c|c}
        \hline
        Model & HR@5 & NDCG@5 & HR@10 & NDCG@10 \\
        \hline
        GPTQ & 0.1020 & 0.0880  &  0.1232 & 0.0951 \\
        SparseGPT & 0.0980 & 0.0830 & 0.1200 & 0.0905  \\
        OD-LLM &  0.0993& 0.0851 & 0.1219 &0.0923 \\
        \hline
    \end{tabular}

    { \textbf{(a) Instruments}}

    \begin{tabular}{c|c|c|c|c}
        \hline
        Model & HR@5 & NDCG@5 & HR@10 & NDCG@10 \\
        \hline
        GPTQ & 0.0992& 0.0808 & 0.1230&  0.0884       \\
        SparseGPT & 0.0800& 0.0620 &0.1180  & 0.0720        \\
        OD-LLM & 0.0838 &0.0644& 0.1264 & 0.0748       \\
        \hline
    \end{tabular}

    { \textbf{(b) Games}}

    \begin{tabular}{c|c|c|c|c}
        \hline
        Model & HR@5 & NDCG@5 & HR@10 & NDCG@10 \\
        \hline
        GPTQ & 0.0796 & 0.0554 & 0.1161 & 0.0672 \\
        SparseGPT &0.0707 &0.0511&    0.1201   &0.0677       \\
        OD-LLM & 0.1173 &0.1000 & 0.1434 & 0.1084 \\
        \hline
    \end{tabular}

    { \textbf{(c) Arts}}
\caption{Comparison with Quantization and Pruning on three datasets.}
\vspace{-10pt}
\end{table}
\subsection{Ablation Study}
To analyze the contributions of each of the individual components of \textbf{OD-LLM}, an ablation study is conducted by evaluating three different variants of the framework.
For simplicity, we denote our complete method as \textbf{M}, and the three variants as \textbf{M-L}, which is LC-Rec, \textbf{M-S} which only includes SVD, and \textbf{M-NS} which includes normalization and SVD. We fix the compression ratio at 0.5. 
Here the complete version of our method \textbf{M} still represents the compressed model using a 50\% compression ratio. 
The results of this study are summarized in Figure \ref{figure.2}. The complete method \textbf{M} consistently delivers the highest performance across all datasets and metrics, showing the strong synergy of all modules. The \textbf{M-L} basically has close performance with OD-LLM. The \textbf{M-S} variant performs particularly poorly, underscoring that normalization is the most critical factor in preserving performance at high compression levels. Adding normalization \textbf{M-NS} yields substantial improvements over \textbf{M-S}, indicating that aligning feature distributions before decomposition is essential to reducing information loss. The additional gains from \textbf{M} over \textbf{M-NS} further demonstrate that the progressive update provides complementary benefits beyond compression and normalization. Overall, these results confirm that every component—normalization, SVD, and layer update contributes meaningfully to the final effectiveness.

\subsection{The Impact of the Compression Ratio}
To investigate the correlation between the compression ratio and the model effectiveness, we select the following representative compression ratio values \{0.2, 0.3, 0.4, 0.5, 0.6, 0.7, 0.8\} and repeat the experiments using each of them.
We evaluate OD-LLM for each compression ratio (from strong to mild compression) on all three datasets, with all other evaluation settings fixed, i.e. the same prompt template, decoding configuration (beam size and maximum generation length), batch size, and data splits.
Importantly, these results use only singular-value decomposition with layer-wise normalization. So they reflect the one-shot behavior of the pipeline. 
The results are shown in Figure \ref{figure.3}. 
The trends show an expected accuracy–compression trade-off: very aggressive compression leads to noticeably lower effectiveness, moderate compression preserves the majority of ranking quality.

\subsection{Comparison with Quantization and Pruning}
We compare our method with two other commonly used model compression techniques: quantization and pruning. 
To ensure a fair comparison, we choose two state-of-the-art post-training models, i.e., GPTQ \cite{frantar2022gptq} and SparseGPT \cite{frantar2023sparsegpt}. 
GPTQ performs layer-wise second-order, weight-only quantization with an error-compensation solver, and SparseGPT applies unstructured pruning guided by a Hessian-aware criterion. 
We align the stored model size for each method: GPTQ uses 4-bit weight-only quantization with group size 128, per-channel quantization and act-order enabled, calibrated on 256 sequences; SparseGPT targets 75\% sparsity on Linear layers using the same 256 calibration without a further quantization; OD-LLM keeps a comparable number of effective parameters. 
All methods are evaluated under identical decoding settings (beam size, max tokens, batch size) on the same single-GPU environment. 
Empirically, GPTQ achieves a slightly higher accuracy than OD-LLM, but its inference latency is around 3x slower than OD-LLM; SparseGPT delivers the worst recommendation effectiveness among the three, and is also not efficient during the pruning stage, whereas OD-LLM is fast in both compression and inference, offering competitive quality with substantially better runtime efficiency.

\subsection{Impact of Calibration Set}
\begin{figure}[t]
    \centering
    \includegraphics[width=0.5\textwidth]{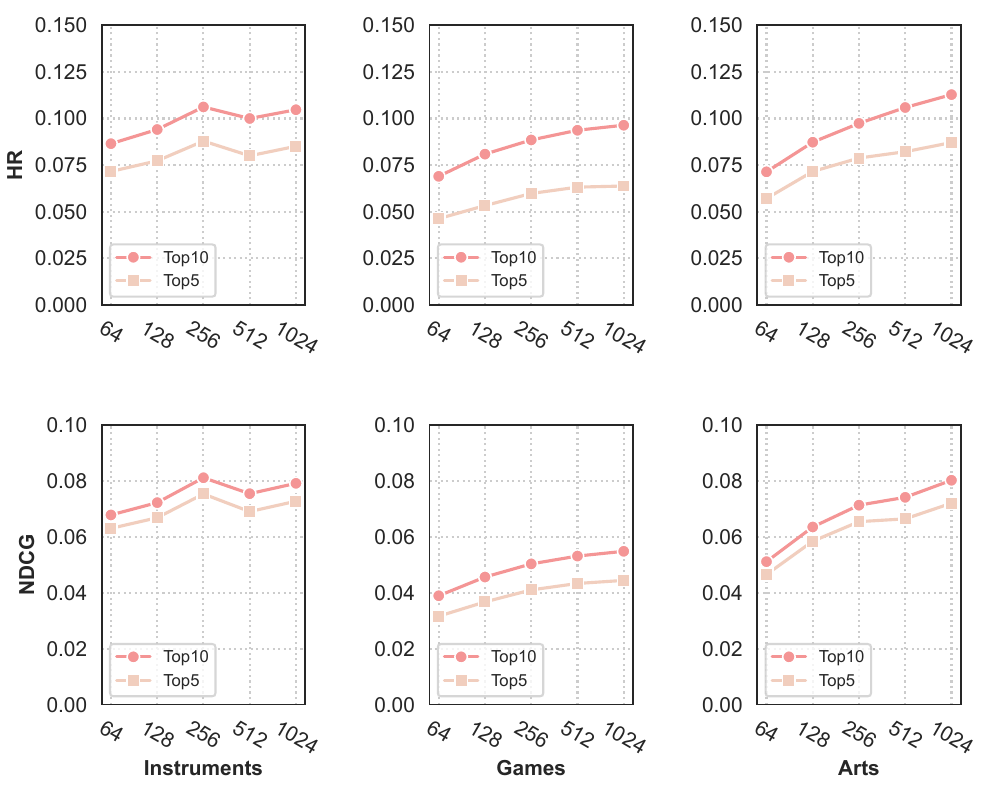}
    \caption{Impact of Calibration Set.}
    \label{figure.4}
    \vspace{-10pt}
\end{figure}

We study the effect of calibration set size by varying the number of calibration examples (64, 128, 256, 512, 1024), while keeping the compression ratio fixed at 0.5, and disabling progressive update -- i.e., a one-shot SVD with layer-wise normalization and no layer update. 
All decoding settings are constant across all runs. The results are in Figure \ref{figure.4}.
The curves show a clear and consistent pattern on the \textit{Instruments}, \textit{Games}, and \textit{Arts} datasets: enlarging the calibration set steadily improves both HR and NDCG for Top-5 and Top-10 sequential recommendation, with the largest gains appearing between the very small sets and the moderately sized ones. 
Once a reasonable number of items is in the set, only marginal improvement gains are achieved, indicating that there are diminishing returns once the calibration distribution has sufficient coverage. 
The separation between the Top-10 and the Top-5 results remain stable across all sizes, and the overall trend is most pronounced on the Arts dataset, suggesting that richer item diversity benefits more from additional calibration.
In practice, this benefit is a trade-off with cost: increasing the calibration set also increases the compression time due to longer activation collection. 
So a moderate calibration budget offers a the best balance between efficiency and effectiveness under this non-progressive setting.

\subsection{Inference Speed Comparison}
Table \ref{Table.4} presents the inference speed comparison of three models: GPTQ, SparseGPT, and OD-LLM—under identical experimental settings, using the same GPU and CPU environments. All models have similar parameter sizes and correspond to those evaluated in Section 4.5. The reported times measure the duration to process one batch of recommendation prompts which is ten prompts. On GPU, OD-LLM achieves a substantial acceleration, requiring only 5 seconds per batch, which is 3.4x faster than GPTQ and 2.4x faster than SparseGPT. On CPU, although all models experience a significant slowdown compared to GPU execution—primarily due to the lack of highly parallel tensor cores, reduced memory bandwidth, and the absence of optimized mixed-precision matrix multiplication kernels—OD-LLM still maintains a clear advantage, running 3.5x faster than GPTQ. This faster execution not only improves latency but also reduces total computation time and energy consumption, making OD-LLM more practical for on-device or edge deployments where GPUs may be unavailable.

\begin{table}[ht]

		\begin{center}
			\begin{tabular}{c|c|c|c}
				\hline
				\ time (s)/batch& GPTQ&SparseGPT&OD-LLM       \\ \hline	
				  GPU & 17 &12 & 5\\
				   CPU&700& 620& 200\\
				\hline
			\end{tabular}
		\end{center}
		\caption{Inference Speed Comparison (Time(s) per batch).}	
		\label{Table.4}
		\vspace{-2em}
	\end{table}
\section{Conclusion}
In summary, this paper proposes OD-LLM, a principled and efficient compression framework designed to enable large language models to operate effectively within the constraints of on-device recommendation systems.
By integrating token-level normalization, low-rank decomposition, and progressive alignment into a unified architecture, OD-LLM delivers substantially smaller models that preserve the nuanced sequential patterns essential to high-quality sequential recommendations. 
Unlike conventional compression methods, OD-LLM is tailored to the behavioral sensitivity and temporal dependencies characteristic to user–item interactions, offering both theoretical rigor and practical viability. Unlike general model compression techniques, our method is much fast for both compression and inference and can be flexibly applied to diffferent LLMs.
Our findings demonstrate that it is possible to retain the expressive power of LLMs for the strict resource demands of edge environments -- a significant step toward privacy-preserving, real-time personalized sequential recommendation on local devices.
\section{Acknowledgments}
This work was partially supported by the Australian Research Council’s Discovery Projects Scheme (DP220101434).

\bibliographystyle{ACM-Reference-Format}
\bibliography{ref}










\end{document}